\documentstyle[amssymb,thmsa,sw20lart]{article}

\input{tcilatex}
\begin{document}

\title{Probability and entropy in quantum theory\thanks{%
Presented at the Eighteenth International Workshop on Maximum Entropy and
Bayesian Methods, MaxEnt'98 (July 27-31, 1998, Garching, Germany).}}
\author{Ariel Caticha \\
{\small Department of Physics, University at Albany-SUNY, Albany, NY 12222%
\thanks{%
E-mail: ariel@cnsvax.albany.edu}}}
\date{}
\maketitle

\begin{abstract}
Entropic arguments are shown to play a central role in the foundations of
quantum theory. We prove that probabilities are given by the modulus squared
of wave functions, and that the time evolution of states is linear and also
unitary.
\end{abstract}

\section{Introduction}

One of the curious features of quantum mechanics%
\index{quantum theory} is that it is a theory in which probabilities play a
most central role and yet, from a foundational point of view, the concept of
entropy is conspicuously absent. Entropy appears only later as an auxiliary
quantity to be used only when a problem is sufficiently complicated that
clean deductive methods have failed and one is forced to use dirtier
inference methods. This is curious indeed because once the use of the notion
of probability has been accepted, the issue of whether or not quantum
mechanics is a theory of inference has been unequivocally settled. Quantum
theory should be regarded as a set of rules for reasoning in situations
where even under optimal conditions the information available to predict the
outcome of an experiment may still turn out to be insufficient. In such a
theory entropy, as the measure of the amount of information \cite{Shannon48}%
\cite{Jaynes83}, should play a central role. It is difficult to avoid the
feeling that perhaps the use of entropic arguments has been inadvertently
encoded into the usual postulates of quantum mechanics. The purpose of this
paper is to show that this is in fact the case.

This paper is a continuation of previous work \cite{Caticha98a}\cite
{Caticha98b} in which quantum theory is formulated as the only consistent
way to manipulate the amplitudes for quantum processes. The result of this
consistent-amplitude approach is the standard quantum theory \cite{Dirac58},
in a form that is very close to Feynman's \cite{Feynman48}. The first and
most crucial step is a decision about the subject matter. We choose a
pragmatic, operational approach: statements about a system are identified
with those experimental setups designed to test them; the objective is to
predict the outcomes of experiments \cite{Stapp72}. The observation that if
two setups are related in some way then information about one may be
relevant to predictions about the other leads one to identify the possible
relations among setups and to introduce the means for handling these
relations quantitatively. This is the role played by the amplitudes.
Amplitudes are tools for reasoning that encode information about how one
builds complicated setups by combining more elementary ones. The question of
how amplitudes are used to predict the outcomes of experiments is addressed
through a single general interpretative rule. A brief summary is given in
section 2.

It is quite remarkable that although the interpretative rule does not in
itself involve probabilities it can be used to prove the Born statistical
``postulate'' (or, better, Born ``rule'') \cite{Finkelstein63} provided one
extra ingredient is added. The point is that the application of the
interpretative rule requires a criterion to quantify the change in
amplitudes when setups are modified. In ref. \cite{Caticha98b} the criterion
adopted was to use the Hilbert norm as the means to measure the distance
between wave functions. This may seem reasonable but such a very technical
assumption without an obvious physical basis clearly detracts from the
beauty of the argument. In section 3 this blemish is corrected. The
realization that the components out of which setups are built, the filters,
already supply us with a notion of orthogonality brings us very close to the
inner product needed for the Hilbert norm. We close this remaining gap using
a symmetry argument. The implication is very interesting: the opinion that
quantum probabilities differ in an essential way from ordinary classical
probabilities is very widespread; not only are they calculated using
apparently different rules but, being given a priori by the Born rule, they
also seem to depend less on information, to be more ``objective''. Our proof
of the Born rule, tracing it to a form of the principle of insufficient
reason, supports the opposite point of view, that such differences are an
illusion, that there is only one kind of probability.

The constraint that amplitudes be assigned consistently leads to a time
evolution that is linear \cite{Caticha98a} but remains silent about whether
it should also be unitary. A common explanation is that imposing unitary
time evolution guarantees that probabilities be conserved. This is true but
it is also irrelevant; that probabilities should add up to one is a matter
of definition \cite{Cox46}; the ``non-conservation of probabilities'' can
always be ``fixed'' by a suitable reinterpretation. The usual symmetry
arguments based on Wigner's theorem are also inadequate. Why should time
evolution be a symmetry in the very technical sense of preserving inner
products? Finally, arguments based on von Neumann's entropy are circular 
\cite{Weinberg89}, they implicitly assume that one can measure observables
other than position \cite{Blankenbecler85}, an assumption which itself (see
section 5) relies on unitary evolution.

The argument we offer in section 4 is based on the idea of array entropy, a
concept that was briefly introduced by Jaynes \cite{Jaynes57} only to be
dismissed as an inadequate candidate for the entropy of a quantum system, a
quantity which he rightfully identified with von Neumann's entropy. From our
point of view, however, amplitudes and wave functions are assigned not just
to the system but to the whole experimental setup, and this turns the array
entropy into a very useful notion. The idea is simple. In situations where
the information available for the prediction of experimental outcomes is not
spoiled by just waiting entropy should be conserved. Our problem is to
identify the appropriate entropy (it is the array entropy). Its conservation
implies the conservation of the Hilbert norm and unitary evolution. As
claimed above, the postulate that time evolution is unitary is derivable
from an entropic argument.

\section{The consistent-amplitude approach to quantum theory}

The objective of quantum theory is to predict the outcomes of experiments;
statements about the quantum system are identified with the experimental
setups designed to test them \cite{Caticha98a}\cite{Caticha98b}. To avoid
irrelevant technical distractions we consider a very simple system, a
particle that lives on a discrete lattice and has no spin or other internal
structure. The generalization to more complex configuration spaces should be
straightforward.

The simplest experimental setup, denoted by $[x_f,x_i]$, consists of placing
a source that prepares the particle at a space-time point $x_i=(%
\vec{x}_i,t_i)$ and placing a detector at $x_f=(\vec{x}_f,t_f)$. To test a
more complex statement such as ``the particle goes from $x_i$ to $x_1$ and
from there to $x_f$,'' denoted by $[x_f,x_1,x_i]$, requires a more complex
setup involving an idealized device, a ``filter'' which prevents any motion
from $x_i$ to $x_f$ except via the intermediate point $x_1$. This filter is
some sort of obstacle or screen that exists only at time $t_1$, blocking the
particle everywhere in space except for a small ``hole'' around $\vec{x}_1$.
The possibility of introducing many filters each with many holes leads to
allowed setups of the general form $a=[x_f,s_N,s_{N-1},\ldots ,s_2,s_1,x_i]$
where $s_n=(x_n,x_n^{\prime },x_n^{\prime \prime },\ldots )$ is a filter at
time $t_n$, intermediate between $t_i$ and $t_f$, with holes at $\vec{x}_n,%
\vec{x}_n^{\prime },\vec{x}_n^{\prime \prime },\ldots $

There are two basic kinds of relations among setups. The first, called $and$%
, arises when two setups $a$ and $b$ are placed in immediate succession
resulting in a third setup which we denote by $ab$. It is necessary that the
destination point of the earlier setup coincide with the source point of the
later one, otherwise the combined $ab$ is not allowed. The second relation,
called $or$, arises from the possibility of opening additional holes in any
given filter. Specifically, when (and {\em only} when) two setups $a^{\prime
}$ and $a^{\prime \prime }$ are identical except on one single filter where
none of the holes of $a^{\prime }$ overlap any of the holes of $a^{\prime
\prime }$, then we may form a third setup $a$, denoted by $a^{\prime }\vee
a^{\prime \prime }$, which includes the holes of both $a^{\prime }$ and $%
a^{\prime \prime }$. Provided the relevant setups are allowed the basic
properties of $and$ and $or$ are quite obvious: $or$ is commutative, but $%
and $ is not; both $and$ and $or$ are associative, and finally, $and$
distributes over $or$.\footnote{%
These are physical rather than logical connectives. They represent our
idealized ability to construct more complex setups out of simpler ones \cite
{Caticha98b} and they differ substantially from their Boolean and quantum
logic counterparts. In Boolean logic not only $and$ distributes over $or$
but $or$ also distributes over $and$ while in quantum logic propositions
refer to quantum properties at one time rather than to processes in time.}

A quantitative representation of $and/or$ is obtained by assigning a single
complex number $\phi (a)$ to each setup $a$ in such a way that relations
among setups translate into relations among the corresponding complex
numbers. What gives the theory its robustness, its uniqueness, is the
requirement that the assignment be consistent: if there are two different
ways to compute $\phi (a)$ the two answers must agree. The remarkable
consequence of the consistency constraints is the possibility of
regraduating $\phi (a)$ with a function $\psi $ to switch to an equivalent
and particularly convenient representation, $\psi (a)\equiv \psi (\phi (a))$%
, in which $and$ and $or$ are respectively represented by multiplication and
addition. Explicitly, $\psi \left( ab\right) =\psi \left( a\right) \psi
\left( b\right) $ and $\psi \left( a\vee a^{\prime }\right) =\psi \left(
a\right) +\psi \left( a^{\prime }\right) $. Complex numbers assigned in this
way are called ``amplitudes''.%
\index{amplitudes} For a similar (earlier) derivation of the quantum sum and
product rules see ref. \cite{Tikochinsky88}.

The observation that a single filter that is totally covered with holes is
equivalent to having no filter at all leads to the fundamental equation of
motion. The idea is expressed by writing the relation among setups $%
[x_f,x_i]=\bigvee_{%
\text{all}\,\,\vec{x}\,\text{at}\,t}([x_f,x_t][x_t,x_i])$ in terms of the
corresponding amplitudes \cite{Feynman48}. Using the sum and product rules,
we get $\psi (x_f,x_i)=\sum_{\text{all}\,\vec{x}\,\text{at}\,t}\psi
(x_f,x_t)\,\psi (x_t,x_i)$.

Following Feynman \cite{Feynman48}, we introduce the wave function%
\index{wave function} $\Psi (%
\vec{x},t)$ as the means to represent those features of the setup prior to $%
t $ that are relevant to time evolution after $t$. Notice that there are
many possible combinations of starting points $x_i$ and of interactions
prior to the time $t$ that will result in identical evolution after time $t$%
. What these different possibilities have in common is that they all lead to 
\hspace{0pt}the same numerical value for the amplitude $\psi (x_t,x_i)$.
Therefore we set $\Psi (\vec{x},t)=\psi (x_t,x_i)$ and all reference to the
irrelevant starting point $x_i$ can be omitted. The traditional language is
that $\Psi $ describes the state of the particle at time $t$, that the
effect of the interactions was to prepare the particle in state $\Psi $. Now
we see that the word ``state'' just refers to a concise means of encoding
information about those aspects of the setup prior to the time $t$ that are
relevant for evolution into the future.

The equation of motion can then be written as

\begin{equation}
\Psi (\vec{x}_f,t_f)=\sum_{\text{all }\,\vec{x}\,\text{at}\,t}\psi (\vec{x}%
_f,t_f;\vec{x},t)\,\Psi (\vec{x},t)\text{,}  \label{motion}
\end{equation}
\newline
which is equivalent to a linear Schr\"{o}dinger equation%
\index{Schrodinger equation} as can easily be seen \cite{Caticha98a}\cite
{Caticha98b} by differentiating with respect to $t_f$ and evaluating at $%
t_f=t$.

This result is important because a variety of nonlinear modifications of
quantum mechanics%
\index{linearity in quantum mechanics} have been proposed, either attempting
to solve the problems with macroscopic quantum superpositions, or to explore
the possibility that the linear theory might just be a low ``intensity''
limit of the true nonlinear theory \cite{deBroglie50}. Furthermore, even
though experimental bounds have become increasingly stringent \cite{Shull80}
experimentation alone cannot logically rule out small nonlinearities.

The question of how amplitudes or wave functions are used to predict the
outcomes of experiments is addressed through the time evolution equation (%
\ref{motion}). For example, suppose the preparation procedure is such that $%
\Psi (%
\vec{x},t)$ vanishes at a certain point $\vec{x}_0$. Then, according to eq. (%
\ref{motion}), placing an obstacle at the single point $(\vec{x}_0,t)$ ({\it %
i.e.}, placing a filter at $t$ with holes everywhere except at $\vec{x}_0$)
should have no effect on the subsequent evolution of $\Psi $. Since
relations among amplitudes are meant to reflect corresponding relations
among setups, it seems natural to assume that the presence or absence of the
filter will have no effect on whether detection at $x_f$ occurs or not.
Therefore when $\Psi (\vec{x}_0,t)=0$ the particle will not be detected at $(%
\vec{x}_0,t)$.

This idea can be generalized to the following general interpretative rule:
Suppose the wave function of a setup is $\Psi \left( t\right) $ and at time $%
t$ one introduces or removes a filter that blocks out those components of $%
\Psi $ characterized by a certain property ${\cal P}$. Suppose further that
this modification of the setup has a negligible effect on the evolution of $%
\Psi $ after $t$. Then when the wave function is $\Psi \left( t\right) $
property ${\cal P}$ will not be detected.

In ref. \cite{Caticha98b} we showed how this interpretative rule implies the
Born statistical postulate provided one uses the Hilbert norm as the means
to quantify the change in the wave function as it evolves through a filter.
In the next section we show why the choice of the Hilbert norm is the
natural one.

\section{The Hilbert inner product}

In order to justify the use of the Hilbert norm%
\index{norm, Hilbert} we show how the concepts of distance and angle among
states, that is an inner product,%
\index{inner product, Hilbert} can be physically motivated. The argument has
three parts.

First, we note that wave functions form a linear space. To illustrate this
point suppose that $\Psi _1(%
\vec{x},t)=\psi (\vec{x},t;\vec{x}_1,t_0)$ is the wave function at time $t$
of a particle that at time $t_0$ was prepared at the point $\vec{x}_1$, and $%
\Psi _2(\vec{x},t)=\psi (\vec{x},t;\vec{x}_2,t_0)$ is the wave function at
time $t$ of a particle that at time $t_0$ was prepared at the point $\vec{x}%
_2$. One way to prepare linear superpositions of $\Psi _1(\vec{x},t)$ and $%
\Psi _2(\vec{x},t)$ is by placing the source at an initial point $(\vec{x}%
_i,t_i)$ with $t_i$ earlier than $t_0$ and letting the particle evolve
through a filter at $t_0$ with holes at $\vec{x}_1$ and $\vec{x}_2$. Then
the amplitude $\psi (\vec{x},t;\vec{x}_i,t_i)$ is 
\begin{equation}
\psi (\vec{x},t;\vec{x}_i,t_i)=\psi (\vec{x},t;\vec{x}_1,t_0)\psi (\vec{x}%
_1,t_0;\vec{x}_i,t_i)+\psi (\vec{x},t;\vec{x}_1,t_0)\psi (\vec{x}_1,t_0;\vec{%
x}_i,t_i)\text{ ,}
\end{equation}
and, in an obvious notation, the wave function at time $t$ is given by the
superposition $\Psi (\vec{x},t)=\alpha \Psi _1(\vec{x},t)+\beta \Psi _2(\vec{%
x},t)$. Notice that the complex numbers $\alpha $ and $\beta $ can be
changed at will by changing the starting point $(\vec{x}_i,t_i)$ or by
modifying the setup between $t_i$ and $t_0$ in any arbitrary way.

The second part of the argument is to point out that the basic components of
setups, the filters, already supply us, without any additional assumptions,
with a concept of orthogonality. The action of a filter $P$ at time $t$ with
holes at a set of points $\vec{x}_p$ is to turn the wave function $\Psi (%
\vec{x})$ into the wave function $P\Psi (\vec{x})=\sum_p\delta _{\vec{x},%
\vec{x}_p}\Psi (\vec{x})$, and since filters $P$ act as projectors, $P^2=P$,
any given filter defines two special classes of wave functions. One is the
subspace of those wave functions such as $\Psi _P\equiv P\Psi $ that are
unaffected by the filter, $P\Psi _P=\Psi _P$. The other is the subspace of
those that are totally blocked by the filter, such as $\Psi _{1-P}\equiv
(1-P)\Psi $, for which $P\Psi _{1-P}=0$. We will say that these two
subspaces are orthogonal to each other.

Any wave function can be decomposed into orthogonal components, $\Psi =\Psi
_P+\Psi _{1-P}$. A particularly convenient expansion in orthogonal
components is that defined by a complete set of elementary filters. A filter 
$P_i$ is elementary if it has a single hole at $\vec{x}_i$, it acts by
multiplying $\Psi (\vec{x})$ by $\delta _{\vec{x},\vec{x}_i}$; the set is
complete if $\sum_iP_i=1$. Then $\Psi (\vec{x})=\sum_iA_i\,\delta _{\vec{x},%
\vec{x}_i}$, where $A_i=\Psi (\vec{x}_i)$.

In the third and last step of our argument, as a matter of convenience, we
switch to the familiar Dirac notation. Instead of writing $\Psi (\vec{x})$
and $\delta _{\vec{x},\vec{x}_i}$ we shall write $|\Psi \rangle $ and $%
|i\rangle $, so that $|\Psi \rangle =\sum_iA_i|i\rangle $. The question is
what else, in addition to the notion of orthogonality described above, is
needed to determine a unique inner product. Recall that an inner product
satisfies three conditions: (a) $\langle \Psi |\Psi \rangle \geqslant 0$
with $\langle \Psi |\Psi \rangle =0$ if and only if $|\Psi \rangle =0$, (b)
linearity in the second factor $\langle \Phi |\alpha _1\Psi _1+\alpha _2\Psi
_2\rangle =\alpha _1\langle \Phi |\Psi _1\rangle +\alpha _2\langle \Phi
|\Psi _2\rangle $, and (c) antilinearity in the first factor, $\langle \Phi
|\Psi \rangle =\langle \Psi |\Phi \rangle ^{*}$. Conditions (b) and (c)
determine the product of the state $|\Phi \rangle =\sum_jB_j|j\rangle $ with 
$|\Psi \rangle =\sum_iA_i|i\rangle $ in terms of the product of $|j\rangle $
with $|i\rangle $, $\langle \Phi |\Psi \rangle =\sum_iB_j^{*}A_i\langle
j|i\rangle $. The orthogonality of the basis functions $\delta _{\vec{x},%
\vec{x}_i}$ is encoded into the inner product by setting $\langle j|i\rangle
=0$ for $i\neq j$, but the case $i=j$ remains undetermined, constrained only
by condition (a) to be real and positive. Clearly an additional ingredient
is needed. What could be more natural than the symmetry argument that if
space itself is homogeneous then there is a priori no reason to favor one
location over another? We therefore choose $\langle i|i\rangle $ equal to a
constant which, without losing generality, we set equal to one. Thus {\em %
the principle of insufficient reason enters quantum theory through the inner
product},%
\index{Principle of Insufficient Reason} $\langle i|j\rangle =\delta
_{ij}\Rightarrow \langle \Phi |\Psi \rangle =\sum_iB_i^{*}A_i$, and this
leads to the Hilbert norm $\left\| \Psi \right\| ^2=\sum_i|A_i|^2$.

One should emphasize that the symmetry argument invoked here differs from
the usual symmetry arguments leading to conservation laws through Noether's
theorem. The latter depends strongly on the particular form of the
Hamiltonian; the former does not.

The deduction of the Born statistical rule now proceeds as in ref. \cite
{Caticha98b}. Briefly the idea is as follows. We want to predict the outcome
of an experiment in which a detector is placed at a certain $%
\vec{x}_k$ when the system is in state $|\Psi \rangle =\sum_iA_i|i\rangle $.
In \cite{Caticha98b} we showed that the state for an ensemble of $N$
identically prepared, independent replicas of our particle is the product $%
|\Psi _N\rangle =\prod_{\alpha =1}^N|\Psi _\alpha \rangle $. Now we apply
the interpretative rule. Suppose that in the $N$-particle configuration
space we place a special filter, denoted by $P_{f,\varepsilon }^k$, the
action of which is to block all components of $|\Psi _N\rangle $ except
those for which the fraction $n/N$ of replicas at $\vec{x}_k$ lies in the
range from $f-\varepsilon $ to $f+\varepsilon $. The difference between the
states $P_{f,\varepsilon }^k|\Psi _N\rangle $ and $|\Psi _N\rangle $ is
measured by the relative Hilbert distance, $||P_{f,\varepsilon }^k|\Psi
_N\rangle -|\Psi _N\rangle ||^2/\langle \Psi _N|\Psi _N\rangle $. The result
of this calculation is \cite{Caticha98b} \ 
\begin{equation}
\stackunder{N\rightarrow \infty }{lim}\,\left\| P_{f,\varepsilon }^k\Psi
_N-\Psi _N\right\| ^2=1-\int_{f-\varepsilon }^{f+\varepsilon }\delta \left(
f^{\prime }-|A_k|^2\right) df^{\prime }\text{ ,}
\end{equation}
where we have normalized $\langle \Psi |\Psi \rangle =\langle \Psi _N|\Psi
_N\rangle =1$. We see that for large $N$ the filter $P_{f,\varepsilon }^k$
has a negligible effect on the state $|\Psi _N\rangle $ provided $f$ lies in
a range $2\varepsilon $ about $|A_k|^2$. Therefore the state $|\Psi
_N\rangle $ does not contain any fractions outside this range. On choosing
stricter filters with $\varepsilon \rightarrow 0$ we conclude that detection
at $\vec{x}_k$ will certainly occur for a fraction $|A_k|^2$ and that it
will not occur for a fraction $1-|A_k|^2$. For any one of the {\em identical}
individual replicas there is no such certainty; the best one can do is to
say that detection will occur with a certain probability $\Pr (k)$. In order
to be consistent with the law of large numbers the assigned value must agree
with the Born rule%
\index{quantum probability},%
\index{Born statistical postulate} 
\begin{equation}
\Pr (k)=|A_k|^2\,.  \label{Bornrule}
\end{equation}

Had we weighted the $|i\rangle $'s differently and chosen a different
normalization $\langle i|i\rangle =w_i$, the probability would be given by $%
\Pr (i)$ $=w_i|A_i|^2$ rather than by eq.(\ref{Bornrule}). It is instructive
to explore this issue further particularly in the continuum limit. Let us
weight each cell of the discrete lattice by its own volume, call it $%
g_i^{1/2}\Delta x$, and let $\Delta x\rightarrow dx$. Replacing $%
w_i^{-1}|i\rangle =(g_i^{1/2}\Delta x)^{-1}|i\rangle $ by$\,|%
\vec{x}\rangle $ the completeness condition $1=\sum_iP_i=\sum_iw_i^{-1}|i%
\rangle \langle i|$ becomes $1=\int g^{1/2}dx\,|\vec{x}\rangle \langle \vec{x%
}|$. Next, replace $\delta _{ij}/\Delta x$ by $\delta (\vec{x}-\vec{x}%
^{\prime })$ and the inner product $\langle i|j\rangle =w_i\delta _{ij}$
becomes $\langle \vec{x}|\vec{x}^{\prime }\rangle =g^{-1/2}\delta (\vec{x}-%
\vec{x}^{\prime })$. Finally, replace $A_i$ by $A(\vec{x})$ and the state $%
|\Psi \rangle =\sum_iA_i|i\rangle $ becomes $|\Psi \rangle =\int
g^{1/2}dx\,A(\vec{x})\,|\vec{x}\rangle $. The Born rule, eq. (\ref{Bornrule}%
), becomes $\Pr (dx)=g^{1/2}dx\,|A(\vec{x})|^2$; $|A(\vec{x})|^2$ is the
probability {\em density}. These results apply both to situations in which
the choice of coordinates is such that the homogeneity of space is not
obvious, and also to intrinsically inhomogeneous, curved spaces.

It is sometimes argued that while there is an element of subjectivity in the
nature of classical probabilities that quantum probabilities are different,
that they are totally objective because they are given by $|A|^2$. We have
just shown that this assignment is neither more nor less subjective than
say, assigning probabilities to each face of a die. Just like one assigns
probability $1/6$ when there is no reason to favor one face of a die over
another, the Born rule follows, even in curved spaces, from giving the same
a priori weight, the same preference, to spatial volume elements that are
equal. (Perhaps it should be the other way around: equally preferred spatial
regions are {\em assigned} equal volumes. This would {\em explain} what a
physical volume is: just a measure of a priori preference.) We have thus
uncovered an interesting connection between quantum theory and the geometry
of space. The full implications of this connection remain to be explored.

\section{Array entropy and unitary time evolution}

When we know everything that is relevant about the experimental setup prior
to time $t=0$ we know $\Psi (\vec{x},0)$; this situation is one of optimal
information. But if less information is available perhaps the best we can do
is conclude that the actual preparation procedure is one among several
possibilities $\alpha =1,2,3,...$ each one with probability $p_{\alpha {}}$.
(For simplicity we initially assume these possibilities form a discrete
set.) The usual linguistic trap is to say {\em the system} is in state $\Psi
_{\alpha {}}(\vec{x},0)$ with probability $p_{\alpha {}}$, but it is better
to say that {\em the preparation procedure} is $\Psi _{\alpha {}}(\vec{x},0)$
with probability $p_{\alpha {}}$. To this state of knowledge, which one may
represent as a set of weighted points in Hilbert space, and which Jaynes
referred to as an array\footnote{%
If the states are normalized the points of the array lie on the surface of a
unit sphere, but normalization is not necessary for our argument.} \cite
{Jaynes57}, one may associate the entropy%
\index{array entropy} 
\begin{equation}
S_A=-\sum_{\alpha {}}p_{\alpha {}}\log p_{\alpha {}}%
\text{ .}  \label{arrayentropy1}
\end{equation}
Jaynes' objection to using this quantity as the entropy of the quantum
system is that if the $\Psi _{\alpha {}}(\vec{x},0)$ are not orthogonal then
the $p_{\alpha {}}$ are not the probabilities of mutually exclusive events.
When regarded as a property or an attribute of the quantum system the
various $\Psi _{\alpha {}}(\vec{x},0)$ need not, in fact, be mutually
exclusive; if $\langle \Psi _{{}\alpha }|\Psi _{\beta {}}\rangle \neq 0$,
knowing that the system is in $\Psi _{{}\alpha }(\vec{x},0)$ does not
exclude the possibility that it will be found in $\Psi _{{}\beta }(\vec{x}%
,0) $. However, if the $\Psi _{\alpha {}}(\vec{x},0)$ are attributes of the
preparation procedure then they are mutually exclusive because the
preparation devices are macroscopic! $S_A$ is the entropy of the preparation
procedure not the entropy of the quantum system.

The importance of this conceptual point cannot be overemphasized and a more
explicit illustration may clarify it further. Consider a spin $1/2$ particle
prepared either with spin along the $z$ direction or with spin along the $x$
direction. These states are not orthogonal and by looking at the particle
there is no sure way to tell which of the two alternatives holds, and yet
the slightest glimpse at the Stern-Gerlach magnets will reveal which of the
two mutually exclusive orientations was used. One can distinguish
non-orthogonal states by looking at the devices that prepared the system
rather than by looking at the system itself.

Turning to the issue of time evolution, we consider situations where those
parts of the setup after time $0$ are known and no further uncertainty is
introduced. Under these conditions the points of the new array are shifted
from $\Psi _{\alpha {}}(\vec{x},0)$ to $\Psi _{\alpha {}}(\vec{x},t)$ but
their probabilities $p_{\alpha {}}$ and the corresponding array entropy $S_A$
remain unchanged.

So far our uncertainty about the preparation procedure was of a rather
simple nature, it led to a probability distribution defined over a discrete
array. But in general there is no such restriction and we may deal with a
continuous array. The simplest continuous array is one dimensional, a
weighted curve $C$ in Hilbert space. We could consider higher dimensional
arrays but this would unnecessarily obscure the argument. The
reparametrization-invariant entropy of this continuous array is \cite
{Jaynes63} 
\begin{equation}
S_A=-\int_Cd\alpha \,p(\alpha )\,\log \,\frac{p(\alpha )}{\ell (\alpha )}%
\text{ ,}  \label{arrayentropy2}
\end{equation}
where $p(\alpha )d\alpha $ is the probability that the preparation procedure
lies in the interval between $\alpha $ and $\alpha +d\alpha $ and $\ell
(\alpha )d\alpha $ is a measure of the distance in Hilbert space between $%
\Psi _{\alpha {}}(\vec{x},0)$ and $\Psi _{\alpha {}+d\alpha }(\vec{x},0)$.
As discussed in the last section the Hilbert norm is the uniquely natural
choice of distance, thus $\ell (\alpha )d\alpha =\left\| |\Psi _{\alpha
{}+d\alpha }\rangle -|\Psi _{\alpha {}{}}\rangle \right\| $.

The possibility of continuous arrays adds a new twist to our considerations
about time evolution. Again we consider setups for which no further
uncertainty is introduced between times $0$ and $t$. We find that points $%
\Psi _{\alpha {}}(\vec{x},0)$ of the old line array at $t=0$ will move to
points $\Psi _{\alpha {}}(\vec{x},t)$ to form a new line array at time $t$.
Since no information was lost between times $0$ and $t$ we expect that, just
as in the discrete case, the probabilities $p(\alpha )d\alpha $ remain
unchanged and the corresponding array entropy $S_A$ is conserved. But
entropy conservation, 
\begin{equation}
\frac{\partial S_A}{\partial t}=\int_Cd\alpha \,\,\frac{p(\alpha )}{\ell
(\alpha )}\,\frac{\partial \ell (\alpha )}{\partial t}=0\text{ ,}
\end{equation}
should hold for any curve $C$ and any function $p(\alpha )$, therefore $%
\partial \ell (\alpha )/\partial t=0$. Thus the conservation of the array
entropy leads to the conservation of Hilbert space distances. Time evolution
must be unitary; the Hamiltonian must be Hermitian.%
\index{unitary time evolution}

\section{Observables other than position}

We have only discussed the measurement of position. How about other
observables,%
\index{observables} uncertainty relations, and so many other notions that
are central in standard quantum theory? Our brief answer (a more detailed
discussion will appear in \cite{Caticha98c}) is that other observables are
useful concepts in that they facilitate the description of complex
experiments but, from our point of view, they are of only secondary
importance and play no role at the foundational level.

One can effectively build more complex detectors by modifying the setup (by
introducing, {\it e.g.}, magnetic fields or diffraction gratings) just prior
to the final position detection at $x_f$. The skill of an experimentalist
consists of arranging the interactions between time $t$ and the time of
detection $t_f$ in such a way that each state $\Phi _n(%
\vec{x},t)$ of an orthogonal set evolves to a corresponding state $\phi _n(%
\vec{x},t_f)=\delta _{\vec{x},\vec{x}_n}$ which also form an orthogonal set.
Then, when the particle is finally detected at time $t_f$ we say that ``at
time $t_f$ the particle was found at $\vec{x}_n$,'' or alternatively, we
convey the same information by saying, somewhat inappropriately, that ``the
particle was found to be in state $\Phi _n(\vec{x},t)$ at time $t$''.

What this particular complex detector ``measures'' is all observables of the
form $Q=\sum_nf_n|\Phi _n\rangle \langle \Phi _n|$. It is noteworthy that
the eigenvalues $f_n$ need not be real; the observables $Q$ are
diagonalizable, {\em i.e.}, normal ($[Q,Q^{\dagger }]=0$), but not
necessarily Hermitian. Clearly, this notion of observables other than
position can only be introduced after one understands that time evolution
must be unitary.

\section{Final remarks}

For over a century now an enormous effort has been directed towards deriving
the second law of thermodynamics from the laws of mechanics. The successive
realization by Gibbs, by Shannon \cite{Shannon48}, and even more so by
Jaynes \cite{Jaynes83} that the validity of entropic arguments rests on
elements that are foreign to mechanics opens the way to inverting the logic
and deriving the laws of mechanics from those same principles of inference
which lie at the heart of thermodynamics. In fact, according to Jaynes'
beautiful explanation \cite{Jaynes65}, the validity of the second law hinges
on the conservation of Gibbs' or von Neumann's entropies. Could this
conservation also be used to deduce unitary time evolution? No. Such
arguments would be circular because these entropies rely for their very
definition on having singled out certain measures (phase space volumes, and
Hilbert norms respectively) as being privileged and the only reason they are
special is precisely that they are conserved under unitary time evolution.

In this work, however, we have given an argument that singles out the
Hilbert norm without appealing to unitarity; this clears the road to
defining an entropy, the array entropy, from the conservation of which one
can deduce the unitarity of time evolution.

The mystery of why complex numbers are sufficient to encode information
about relations between setups remains. It seems that one could use other
mathematical objects with the required associativity and distributivity, for
example matrices or other Clifford numbers \cite{Finkelstein62}. The recent
work by Rodr\'{\i }guez \cite{Rodriguez98} may contain important steps in
exploring this possibility from a rather different point of view. My own
belief is that the connection between the quantum inner product and spatial
measures of volume strongly suggests that the reason for complex numbers
will be found in the geometry of space. Perhaps eventually even the geometry
of space itself will be determined by entropic considerations as well.

\noindent {\bf Acknowledgments-} I am indebted to C. Rodr\'{\i }guez and P.
Zambianchi for valuable discussions and many insightful remarks.
Correspondence with J. Hartle and L. Schulmann on the issue of the Hilbert
norm is also gratefully acknowledged.

\end{document}